\documentclass[summary]{URSIGASS2021}

\usepackage{amsmath}

\title{An evaluation of the data space dimension in phase retrieval: results in Fresnel zone}

\author{Rocco Pierri\affref{ref1}, Raffaele Moretta\affref{ref1}}

\affiliation{
  \aff{ref1}{Dipartimento di Ingegneria, Universita' della Campania "L. Vanvitelli", Aversa, 81031, Italia.
}
}

\begin{document}

\maketitle

\begin{abstract}
  In this paper, we address the problem of computing the dimension of data space in phase retrieval problem.\\ Starting from the quadratic formulation of the phase retrieval, the analysis is performed in two steps. First, we exploit the \textit{lifting technique} to obtain a linear representation of the data. Later, we evaluate the dimension of data space by computing analytically the number of relevant singular values of the linear operator that represents the data.\\
  The study is done with reference to a $2D$ scalar geometry consisting of an electric current strip whose square amplitude of the electric radiated field is observed on a two-dimensional extended domain in Fresnel zone.
\end{abstract}

\section{Introduction}
\label{sec:introduction}
Phase retrieval techniques find applications in all the contexts where phase information is not available. In electromagnetism, they arise in antenna or array diagnostics, in the reconstruction of the far-field pattern from near-zone data (phaseless near-field far-field techniques) \cite{1,fuchs}, and in inverse scattering problem.\\ From the mathematical point of view, the lack of phase information makes the problem non-linear and this complicates the task of finding a solution. Over the years, different numerical procedures to address the problem have been proposed; some of them exploit the amplitude formulation, instead, others are based on the square amplitude formulation. The latter consists in retrieving the unknown function $f\in{X}$ from the quadratic model below 
\begin{equation}
	|T f|^2=|g|^2
\end{equation}
where $T:f\in X \longrightarrow g\in Y$.\\ 
The most common techniques to tackle the problem exploit a least-square minimization. However, since the cost functional to be minimized is not quadratic, trap points may occur. The latter may avoid reaching the actual solution of the problem even if the uniqueness conditions are satisfied.\\ 			          			        To overcome this drawback, the \textit{lifting technique} can be used \cite{4}. The latter, starting from the quadratic formulation in (1), exploits a redefinition of unknown space to recast the phase retrieval problem as a linear one. Despite this, the new unknown function belongs to a functional space whose dimensions are the square of those of the original unknowns space. Consequently, for large-scale problems the lifting approach is not feasible and the phase retrieval problem must be necessarily addressed by recurring to non-convex formulations.\\
In this framework, avoiding trap points is the main task. From this point of view, the least-squares minimization based on the square amplitude formulation brings to a cost functional that is smoother that obtained by considering amplitude formulation. Furthermore, the quadratic formulation of phase retrieval problem allows a deep analysis of the genesis of local minima and allows finding strategies to ``cure'' them. In particular, it has been shown that if the ratio between the dimension of data space ($M$) and the dimension of unknowns space ($N$) is high enough, no trap points appear in the functional to minimize \cite{2,3,Sun}. From this discussion, it is evident that the dimension of data space plays a key role in phase retrieval via quadratic approach; hence, it is worth investigating how to evaluate it from an analytical point of view.\\
As shown in \cite{4}, the dimension of data space can be evaluated by counting the number of significant singular values of the lifting operator. However, to the best of our knowledge, analytical results concerning the singular values behavior of lifting operator are not available in literature.\\ For such reason, with reference to a $2D$ scalar geometry, we will provide a closed-form expression of the number of significant singular values of the pertinent lifting operator.
%
\section{Geometry of the problem and preliminary results}
In this paper, we consider the $2D$ scalar geometry depicted in fig. 1 where the $y$-axis represents the axis of invariance.\\
An electric current $\underline J(x) = J(x)\,\hat i_y$ supported on the set
$[-a, a]$ of the $x$-axis radiates within a homogeneous medium with wavenumber $\beta$.\\
The electric field $\underline E$ radiated by such strip source has one component directed along the $\hat i_y$; hence, $\underline E(r,\theta)=E(r,\theta)\,\hat i_y $.
The square amplitude of the radiated electric field $|E|^2$ is observed in Fresnel zone on a two-dimensional domain that extends along the polar coordinates $(r,\theta)$ on the set $[r_{min},r_{max}]\times[-\theta_{max},\theta_{max}]$.\\
\noindent For the geometry at hand, the radiated electric field can by expressed in the variables $r$ and $u=sin(\theta)$ by the equation 
\begin{equation}
	E(r,u)=T J(x)
\end{equation} 
where $T$ is the linear integral operator that realizes the following mapping
\begin{equation}
	T : J \in L_2(SD) \longrightarrow  E\in L_4(OD)
\end{equation}
with $L_2(SD)$ denoting the space of square-integrable functions on the set $SD=[-a,a]$, and $L_4(OD)$ indicating the space of functions whose amplitude to the fourth power is integrable on the set $OD=[r_{min},r_{max}]\times[-u_{max},u_{max}]$.\\
\noindent Under the paraxial Fresnel approximation, the operator $T$ can be explicitly written as in \cite{5} in the form
\begin{equation}
	T\,J=\frac{1}{\sqrt{\beta r}}{e^{\,-j\beta r\,(1+\frac{1}{2}u^2)}}\int_{-a}^{+a}e^{-j\frac{\beta}{2r}{x}^2}  e^{\,j\beta u x}\,J(x)dx
\end{equation}
\begin{figure}[t]
	\centering
	\includegraphics[scale=0.65]{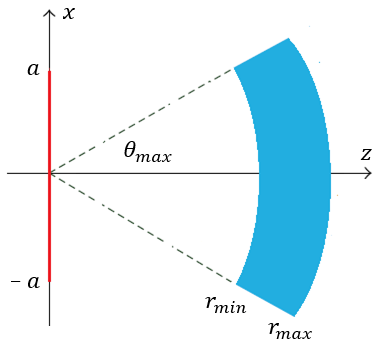}
	\caption{Geometry of the problem}
\end{figure}

\section{The lifting operator and its singular values}
\begin{figure}
	\centering
	\includegraphics[scale=0.125]{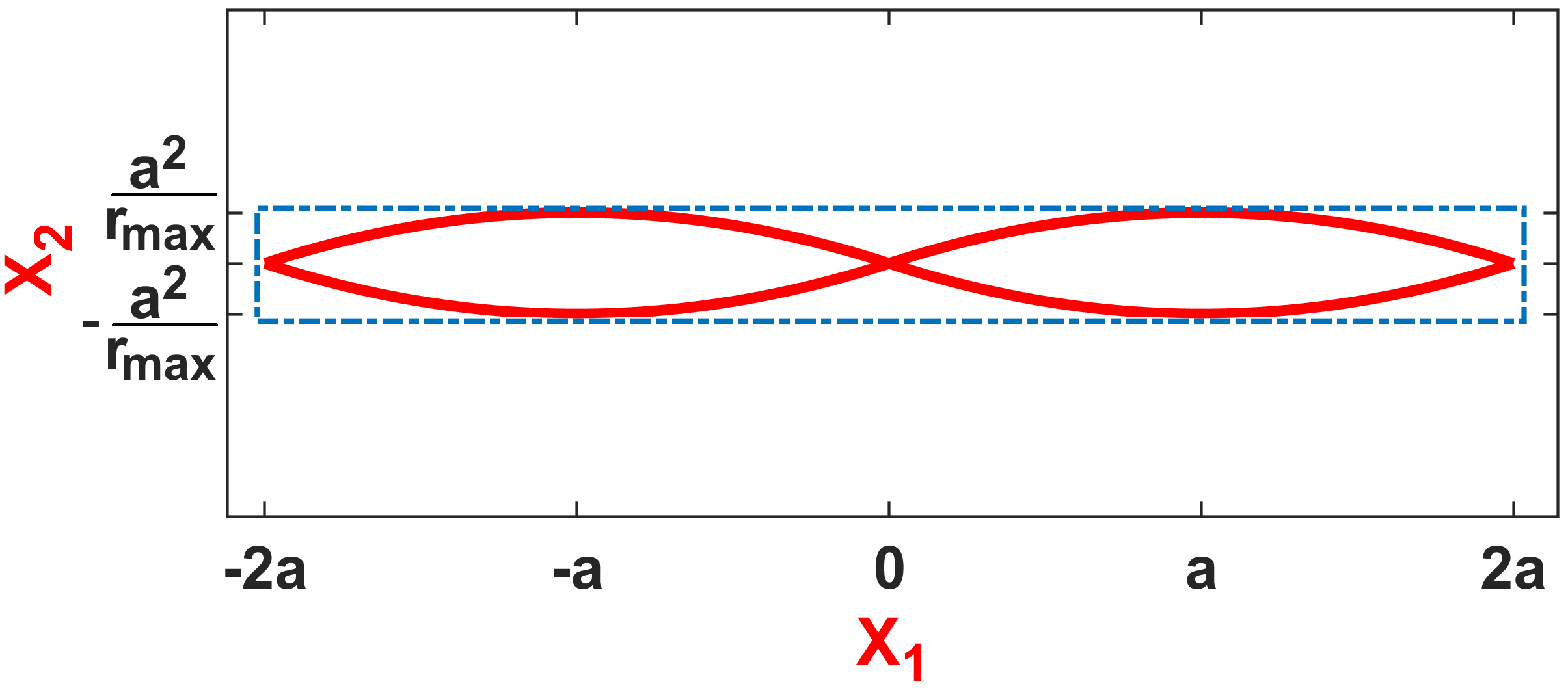}
	\caption{Shape of $\hat D_1$ when $a=10\lambda$, $r_{max}=100\lambda$.}
	\label{fig2}
\end{figure}

In this section, first we provide a linear representation of $|E(r,u)|^2$, i.e, the square amplitude of the electric field over the observation domain $OD$. Later, we find the dimension of data space by evaluating the number of significant singular values of the linear operator which represents the data.\\  
To obtain a linear representation of $|E|^2$, let us rewrite the quadratic model $|E|^2=|T J|^2$ in the form below
$$|E(r,u)|^2=\big(TJ\big)\,\big(TJ\big)^*=$$ 
\begin{equation}
	=\frac{1}{{\beta r}}\iint_{D} e^{\,j\frac{\beta}{2r}({\overline x}^2-{x}^2)}\, e^{-j\beta u (\overline x-x)}\,J(x)\,J^*(\overline x)\,dx\,d\overline x
\end{equation}
with $D=\{(x,\overline x)\in[-a,a]\times[-a,a]\}$.\\
From the last equation, it is evident that if we redefine the unknown space and we consider as unknown the function $F(x,\overline x) = J(x)\,J^*(\overline x)$
 then the operator which links the unknown function $F(x,\overline x)$ with the data function $|E(r,u)|^2$ is linear. Such operator is known in literature as \textit{lifting operator} and it is defined as
\begin{equation}
	A\, : \,  F\in L_{2\,}(SD\times SD)\, \longrightarrow \, |E|^2\in L_2^+(OD)
\end{equation}
where
\begin{equation}
	AF=\frac{1}{{\beta r}}\iint_{D}e^{\,j\frac{\beta}{2r}({\overline x}^2-{x}^2)}\, e^{-j\beta u (\overline x-x)}\,F(x,\overline x)\,dx\,d\overline x
\end{equation}
A weighted adjoint operator $A_w^\dag$ is given by 
\begin{equation}
	A_w^\dag(\cdot) =\int_{r_{min}}^{r_{max}}\int_{-u_{max}}^{+u_{max}} \frac{w(x,\overline x)}{{\beta r}}e^{-j\frac{\beta}{2r}({\overline x}^2-{x}^2)} e^{j\beta u (\overline x-x)}\,(\cdot)\,du\,dr
\end{equation}
where $w(x,\overline x)$ is a weight function, and $(\cdot)$ denotes the function of the variables $(r,u)$ on which the adjoint operator acts.\\
Naturally, the presence of the weight function changes the dynamics of the singular values of the lifting operator $A$. Despite this, we will show that  the \textit{number of relevant singular values} remains unchanged. For such reason, we can tolerate the changes in the singular values behavior brought by the weight function.\\
To evaluate the number of significant singular values of the lifting operator, we will find the eigenvalues $\lambda_m$ of the operator $AA_w^\dag$.
The latter can be expressed as 
\begin{equation}
	AA_w^\dag(\cdot)=\int_{r_{min}}^{r_{max}}\int_{-u_{max}}^{+u_{max}}H(r,r_o,u,u_o)\,(\cdot)\,du\,dr
\end{equation}
where
\begin{equation}
		H(r,r_o,u,u_o)= \hspace{140pt}
\end{equation}
\begin{displaymath}
\hspace{10pt}\frac{1}{\beta^2 rr_o}\iint_{D}w(x,\overline x)\,e^{j\frac{\beta}{2}\left(\frac{1}{r_o}-\frac{1}{r}\right)({\overline x}^2-{x}^2)} e^{-j\beta (u_o-u) (\overline x-x)}dxd\overline x
	\label{kernel_AAagg}
\end{displaymath}
\vspace{1pt}\\
\noindent With the aim to compute the integral (10) in a very simple way, let us divide the integration domain $D$ as
$$D=D_1 \cup D_2$$
where
$D_1=\{(x,\overline x)\in D:x\neq\overline x\}$, $D_2=\{(x,\overline x)\in D:x=\overline x\}$.\\
The set $D_2$ is a null set with respect to the Lebesgue measure; consequently, the kernel can be computed by performing the integration only on $D_1$.\\ 
With the aim to evaluate $H(r,r_o,u,u_o)$, let us perform the change of variables
\begin{equation}
	\left\{\begin{array}{c}
		X_1=\overline x - x\vspace{3pt}\\
		X_2=\dfrac{\overline x^2 - x^2}{r_{max}}
	\end{array} \right.
\label{trasformazione}
\end{equation}
which is injective and continuously differentiable on $D_1$. By virtue of \eqref{trasformazione}, the kernel of $AA_w^\dag$ 
can be recast as
\begin{equation}
		H(r,r_o,u,u_o)=\dfrac{1}{\beta^2 rr_o} \hspace{120pt}
\end{equation}
$$
\iint_{\hat D_1}w(X_1,X_2)e^{j\frac{\beta}{2}\left(\frac{r_{max}}{r_o}-\frac{r_{max}}{r}\right)X_2} e^{-j\beta (u_o-u) X_1} \left|\frac{\partial(x,\overline x)}{\partial(X_1,X_2)}\right|dX_1dX_2
$$
where
\begin{itemize}
	\item  $\left|\frac{\partial(x,\overline x)}{\partial(X_1,X_2)}\right|=-\dfrac{r_{max}}{2X_1}$ denotes the Jacobian determinant of the transformation,
	\item  $\hat D_1$ indicates the domain in which the original integration domain $D_1$ is mapped by \eqref{trasformazione}.
\end{itemize}
Note that, despite the Jacobian determinant is singular for $X_1=0$, such point does not belong to the integration domain $\hat D_1$. 
Hence, no singularity appears in the integral (12).
Now, if we choose
\begin{equation}
	w(X_1,X_2)=\left(\, \left|\frac{\partial(x,\overline x)}{\partial(X_1,X_2)}\right|\, \right)^{-1}
\end{equation}
we have that
$$	H(r,r_o,u,u_o)=$$
\begin{equation}
\dfrac{1}{\beta^2 rr_o}\iint_{\hat D_1}e^{j\frac{\beta }{2}\left(\frac{r_{max}}{r_o}-\frac{r_{max}}{r}\right)X_2} e^{-j\beta (u_o-u) X_1} dX_1\,dX_2
	\label{kernel2}
\end{equation}
\noindent According to equation \eqref{kernel2}, the integration should be done on the set $\hat D_1$ which is sketched in fig. \ref{fig2}. However, since we want to recast the operator $AA_w^\dag$ in a form whose eigenvalues are known in closed-form, we will approximate $H(r,r_o,u,u_o)$ by integrating on the smaller domain that encloses the set $\hat D_1$. The latter is made up by all the points $(X_1,X_2)$ belonging to the rectangular set $[-2a,2a]\times[-{a^2}/{r_{max}},\,{a^2}/{r_{max}}]$ except for the point $(0,0)$.
By performing the integration on such domain, we have that 
%
\begin{equation}
	\begin{split}
	&H(r,r_o,u,u_o)\approx\hspace{50pt}\\&\hspace{0pt}\frac{8a^3}{\beta^2 r_{max}}\dfrac{1}{r r_o}sinc\left(\frac{\beta a^2}{2}\left(\frac{1}{r_o}-\dfrac{1}{r}\right)\right)sinc\left(2\beta a (u_o-u)\right)\vspace{2pt}\\
	\end{split}
\end{equation}
Accordingly, the operator $AA_w^\dag$ can be expressed as 
\begin{equation}
	\begin{split}
		&AA_w^\dag(\cdot)\approx\dfrac{8a^3}{\beta^2r_{max}}\int_{r_{min}}^{r_{max}}\int_{-u_{max}}^{+u_{max}}\frac{1}{r r_o} sinc\left(\frac{\beta a^2}{2}\left(\frac{1}{r_o}-\frac{1}{r}\right)\right)\vspace{12pt}\\& \hspace{123pt} sinc\left(2\beta a \,(u_o-u)\right) \,(\cdot)\,du\,dr
	\end{split}
\label{operatore_esplicito}
\end{equation}
From \eqref{operatore_esplicito}, it is evident that the operator $AA_w^\dag$ becomes more similar to a convolution operator if we set $s=r_{max}/r$. In fact, by doing this, we have that 
\begin{equation}
	\begin{split}
		&AA_w^\dag(\cdot)\approx\dfrac{8a^3}{\beta^2r_{max}^2}\int_{1}^{\frac{r_{max}}{r_{min}}}\int_{-u_{max}}^{+u_{max}}\dfrac{s_o}{s}\,sinc\left(\dfrac{\beta a^2}{2r_{max}}\left(s_o-s\right)\right)\vspace{2pt}\\&\hspace{115pt}sinc\big(2\beta a (u_o-u)\big)\,(\cdot)\,du\,ds
		\label{approx2}
	\end{split}
\end{equation}
Despite the previous operator is only an approximation of $AA_w^\dag$, the positive aspect is that its eigenvalues are known in closed-form.\\  
It is worth noting that the operator $\eqref{approx2}$ has been obtained by exploiting two changes of variables. The idea of exploiting a change of variable in such a way that the kernel of the considered operator assumes a desired form has been exploited also in other recent works \cite{warping1}.\\ 
In order to compute the eigenvalues of \eqref{approx2}, we must solve the eigenvalue problem 
\begin{equation}
AA_w^\dag v_m = \lambda_m v_m
\end{equation}
where $v_m$ represents the $m-th$ eigenfunctions of $AA_w^\dag$.\\
By fixing $\tilde v_m(s,u) = {v_m(s,u)}/{s}$ the eigenvalue problem above can be recast as\\
\begin{equation}
	\begin{split}
		&\dfrac{8a^3}{\beta^2 r_{max}^2}\int_{1}^{\frac{r_{max}}{r_{min}}}\int_{-u_{max}}^{u_{max}}sinc\left(\dfrac{\beta a^2}{2r_{max}}\left(s_o-s\right)\right)sinc\big(2\beta a (u_o-u)\big)\vspace{2pt}\\&\hspace{120pt}\,\tilde{v}_{m}(s,u)\,du\,ds\,=\,\lambda_m v_m(s_o,u_o)
	\end{split}
    \label{eq19}
\end{equation}
The eigenvalues of \eqref{eq19} are known in closed-form. In fact, according to \cite{6}, they are given by the equation
\begin{equation}
	\lambda_m(AA_w^\dag)=\lambda_{m_1}^{(u)}\,\lambda_{m_2}^{(s)}
\end{equation}
where $\lambda_{m_1}^{(u)}$ and $\lambda_{m_2}^{(s)}$ denote the eigenvalues of the Slepian-Pollak operators whose kernels are respectively $sinc\,(\,2\beta a\,(s_o-s)\,)$ and $sinc\,\big(\frac{\beta a^2}{2r_{max}}(s_o-s)\big)$.\\ Since the sequences $\{\lambda_{m_1}^{(u)}\}$ and $\{\lambda_{m_2}^{(s)}\}$ are relevant respectively until to the indexes
{\begin{equation}
		M_u=\dfrac{4}{\pi}\,\beta\,a\,u_{max}+1 \ \,\,\,\,\,\,
		M_s=\dfrac{\beta a^2}{2\pi}\left(\dfrac{1}{r_{min}}-\dfrac{1}{r_{max}}\right)+1,
\end{equation}}
it results that the eigenvalues of the problem (19) 
and, consequently, the eigenvalues of the operator (17) are significant until to the index $\overline{M}=M_u\,M_s$.\\
Let us remember that the kernel of the operator in (17) has been obtained by integrating  on the smaller rectangular that encloses $\hat D_1$; for such reason, $\overline{M}$ is not exactly equal to the number of relevant eigenvalues of $AA_w^\dag$ but it represents an upper bound.\vspace{0pt}\\
Until now, we have focused on the eigenvalues of $AA_w^\dag$; instead, the actual singular values of $A$ are related to the eigenvalues of $AA^\dag$ by the equation
$\sigma_m(A)=\sqrt{\lambda_m(AA^\dag)}$ 
where $A^\dag$ denotes the usual adjoint operator defined without the weight function.
For such reason, we know only an approximation of the singular values of A that is given by the square root of the eigenvalues of $AA_w^\dag$.\\
In the next section, by means of some simulations, we will check that the actual singular values of A and their approximated version become negligible at the same index. This verification allows to state that for the considered geometry,
the number of significant singular values of the lifting operator or in other words the dimension of data space $M$ satisfies the inequality
\begin{equation}
	M\leq M_u\, M_s
\end{equation}

\section{Numerical results}
In this section, we check that actual singular values of $A$ and their approximations become negligible at the same index.
As test case, we consider the configuration in which $a=10\lambda$, $u_{max}=0.5$,
$r_{min}=25\lambda$ ($s_{max}=4$), $r_{max}=100\lambda$ ($s_{min}=1$). With reference to such configuration, in fig. 3 we have sketched the actual singular values of A and their approximated versions in dB. In particular, the blue, red and black diagrams sketch respectively 
\begin{itemize}
	\item the square root of the eigenvalues of the approximated version of $AA_w^\dag$ provided by (17),
	\item the square root of the eigenvalues of $AA^{\dag}_w$,
	\item the square root of the eigenvalues of $AA^\dag$.
\end{itemize}
\begin{figure}[t]
	\centering
	\includegraphics[scale=0.12]{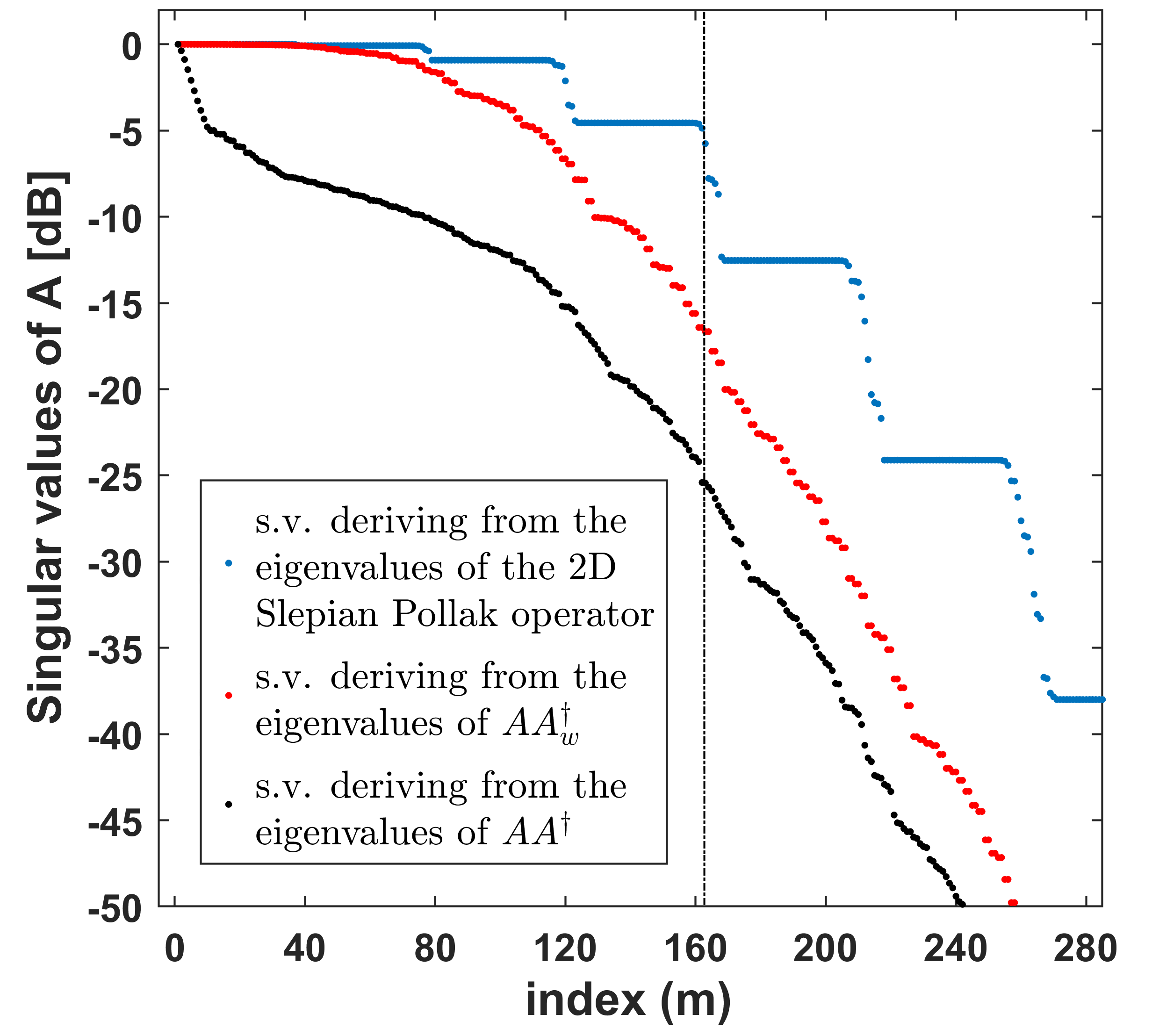}
	\caption{Singular values of $A$, and their approximated versions in dB.}
\end{figure}
As can be seen from fig. 3, the square root of the eigenvalues of the operator (17) exhibits a multi-step behavior and they are relevant until to the index $\overline{M}=M_uM_s=164$.\\
The multi-step behavior can be understood if we remember that the eigenvalues of such an operator are given by (20). Now, as shown in fig. 4,  the sequence $\{\lambda_{m_1}^{(u)}\}$ have a step-like behavior, instead,  the sequence $\{\lambda_{m_2}^{(s)}\}$ is not exactly step-like. This automatically implies that the eigenvalues of the operator (17) have a multi-step behavior also before the index $\overline M =164$ and, consequently, also their square root.\\ 
However, our aim is to forecast the critical index at which the actual singular values of A become negligible. By observing the behavior of the actual singular values (black diagram in fig. 3), it is evident that the singular values beyond the index $\overline M=164$ are surely negligible while those before are almost all significant if the noise level is not so high.
This implies that the use of the weighted adjoint changes only the dynamics of the singular values but not the critical index at which they become negligible. For such reason, we can state that  $\overline M=M_uM_s$ is an upper bound for the dimension of data space that is very close to its actual value.   

\begin{figure}[t]
	\centering
	\includegraphics[scale=0.105]{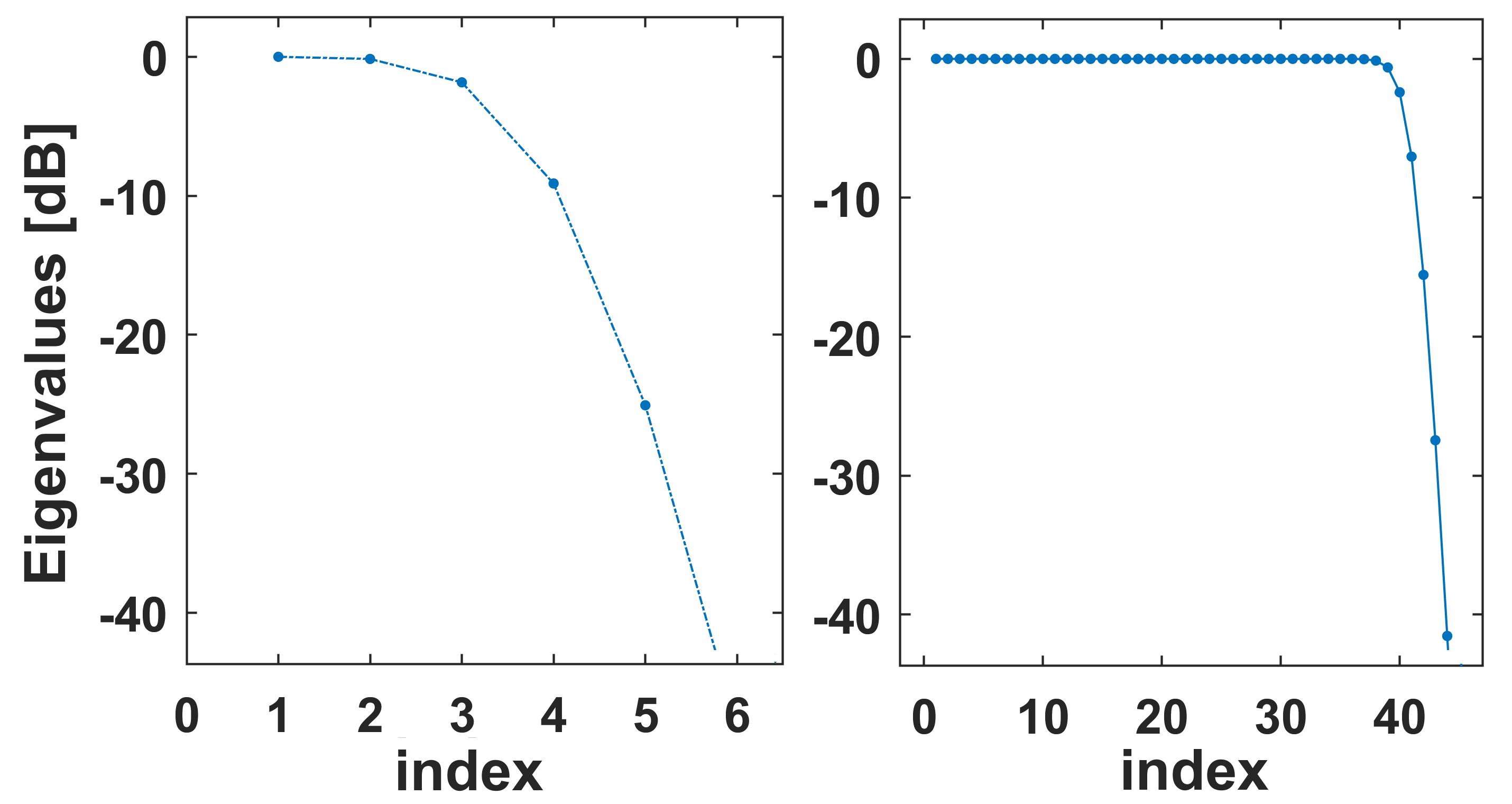}
	\caption{Eigenvalues of the Slepian-Pollak operators whose kernels are $sinc\,(2\beta a \,(u_o-u))$ ({left panel}), $sinc\,\big(\frac{\beta a^2}{2r_{max}}(s_o-s)\big)$ ({right panel}).}
\end{figure}

\section{Conclusion} 
In this article, we have addressed the problem of evaluating the dimension of data space in phase retrieval. In particular, with reference to a $2D$ geometry consisting of a strip current observed on a two dimensional observation domain, we first have introduced a linear operator that represents the square amplitude of the radiated field. After, studying the singular values of such an operator, we have provided an upper bound for the dimension of data space which is very near to its actual value.

\end{document}